\documentclass[%
reprint,
superscriptaddress,
longbibliography,
showpacs,
amsmath,amssymb,
aps,
]{revtex4-1}
\usepackage[normalem]{ulem}
\usepackage{color}
\usepackage{graphicx}
\usepackage{amsthm}
\usepackage{amsmath}	
\usepackage{mathtools}
\usepackage{upgreek}
\usepackage{layouts}

\def\U#1{{%
\def\O{\mbox{O}}
\def\u{\mbox{u}}
\mathcode`\u=\upmu
\mathcode`\O=\Upomega
\mathrm{#1}}}
\def\sub#1{_{\mbox{\scriptsize#1}}}
\def\sur#1{^{\mbox{\scriptsize#1}}}

\def\jj{\,\mathrm{j}}                   % thin space before j

\def\Im{\mathop{\mathrm{Im}}}
\def\vct#1{{\mathchoice{\mbox{\boldmath$#1$}}{\mbox{\boldmath$#1$}}%
  {\mbox{\scriptsize\boldmath$#1$}}{\mbox{\scriptsize\boldmath$#1$}}}}

\begin{document}
\title{
Reconfigurable terahertz quarter-wave plate for
helicity switching \\ based on
Babinet inversion of anisotropic checkerboard metasurface
} 

\author{Yosuke~Nakata}
\email{nakata@qc.rcast.u-tokyo.ac.jp}
\affiliation{Research Center for Advanced Science and Technology, The University of Tokyo, 4-6-1 Komaba, Meguro-ku, Tokyo 153-8904, Japan}
\author{Kai~Fukawa}
\affiliation{Department of Physics, Faculty of Science, Shinshu University, 3-1-1 Asahi, Matsumoto, Nagano 390-8621, Japan}
\author{Toshihiro~Nakanishi}
\affiliation{Department of Electronic Science and Engineering, Kyoto University, Kyoto 615-8510, Japan}
\author{Yoshiro~Urade}
\affiliation{Center for Emergent Matter Science, RIKEN, 2-1 Hirosawa, Wako, Saitama 351-0198 Japan}
\author{Kunio~Okimura}
\affiliation{School of Engineering, Tokai University, 4-1-1 Kitakaname, Hiratsuka, Kanagawa 259-1292, Japan}
\author{Fumiaki~Miyamaru}
\affiliation{Department of Physics, Faculty of Science, Shinshu University, 3-1-1 Asahi, Matsumoto, Nagano 390-8621, Japan}
\affiliation{Center for Energy and Environmental Science, Shinshu University, 4-17-1 Wakasato, Nagano 380-8553, Japan}

\date{Compiled \today }


\begin{abstract}
 Dynamic helicity switching by utilizing metasurfaces is challenging because
 it requires deep modulation of polarization states.
 To realize such helicity switching,
 this paper proposes a dynamic metasurface functioning as a
 switchable quarter-wave plate,
 the fast axis of which can be dynamically rotated by $90^\circ$.
 The device is based on the critical transition of an anisotropic metallic checkerboard,
 which realizes the deep modulation and simultaneous design of the switchable states.
 After verifying the functionality of the ideally designed device in a simulation,
 we tune its structural parameters to realize practical experiments in the terahertz frequency range.
 By evaluating the fabricated sample with vanadium dioxide, the conductivity of which can be controlled by temperature,
 its dynamic helicity switching function is demonstrated.
\end{abstract}

\maketitle
\section{Introduction}
Dynamic polarization control technology has been fundamental 
for flat-panel displays, polarization-sensitive spectroscopy,
highly sensitive measurements with the lock-in detection technique,
and data transmission.
In the optical frequency domain, liquid-crystal and electro-optic-crystal modulators
have been widely used to achieve dynamic polarization modulation.
However, in lower frequency ranges such as the terahertz frequency range \cite{Lee2009},
which is potentially important for spectrally resolving giant and bio-molecules,
the wavelength is several orders of magnitude larger than that in the optical frequency range.
Thus, the thickness of conventional dielectric devices at terahertz frequencies
must become much larger than that at optical frequencies,
and it is difficult to reach subwavelength-order dynamic devices.

To overcome this difficulty, researchers have been developing
artificially designed surfaces, called {\it metasurfaces} \cite{Glybovski2016, Chen2016, Al-Naib2017}.
Most metasurfaces are partially made of metals,
the highly conductive nature of which is
applicable to subwavelength devices.
Recently, the progress of metasurfaces has motivated research into dynamic terahertz polarization control utilizing 
photoswitchable chiral metamolecules \cite{Zhang2012},
deformable MEMS (micro electro mechanical systems) spirals \cite{Kan2013,Kan2015},
gate-tunable polarization modulators with graphene \cite{Miao2015},
tunable MEMS cross-polarization converters \cite{Zhang2017},
and tunable wave plates with actuated microcantilever arrays to switch circular and linear
polarization states of transmitted waves \cite{Zhao2018}.
Phase change materials such as vanadium dioxide have also been applied to 
dynamic quarter-wave plates with variable working frequencies \cite{Wang2015b,Wang2016}
and reconfigurable polarization modulators between linear and circular polarization states \cite{Liu2018, Nouman2018}.
Despite these remarkable achievements, designing switchable metasurfaces
that have desired functionalities with deep modulation contrast is still challenging.
This is because it also requires the simultaneous design to realize multiple functionalities,
which usually interfere with each other.

\begin{figure}[!b]
\centering
 \includegraphics[width=2.4in]{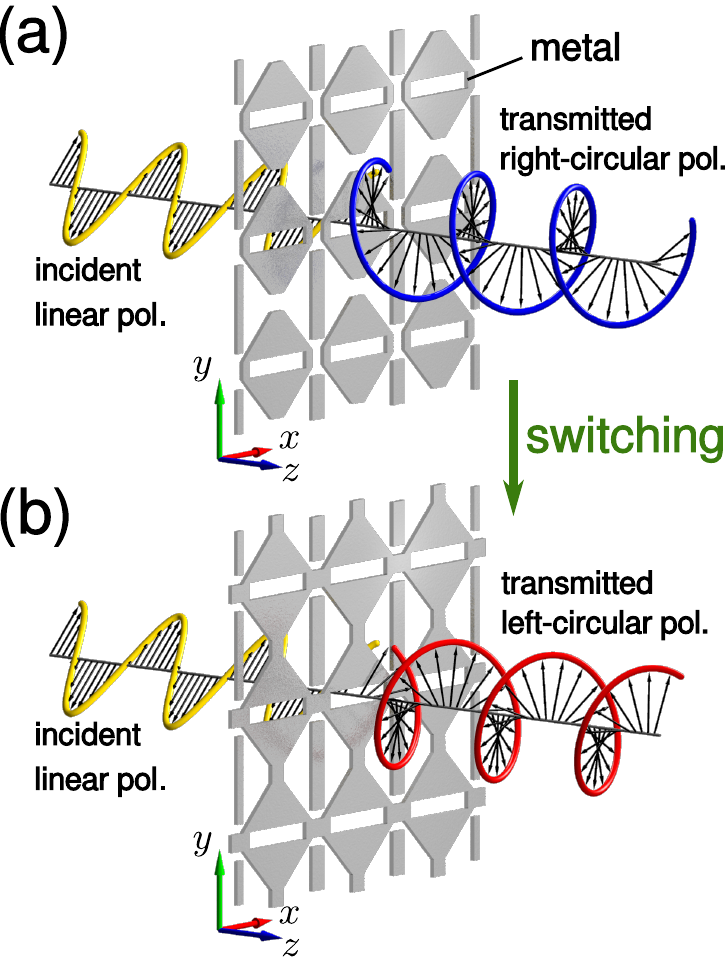}
 \caption{Schematics for dynamic helicity switching for the transmitted
 wave by utilizing the transition between the (a) disconnected and (b) connected anisotropic checkerboards.
 \label{fig1}}
\end{figure}

To resolve this issue, an extremely thin metallic checkerboard can be a key structure.
The electromagnetic properties of metallic checkerboards critically depend on their connection states
\cite{Compton1984, Kempa2010, Edmunds2010, Takano2014, Tremain2015}.
Between the connected and disconnected states, extraordinary frequency-independent characteristics have been revealed for the metallic checkerboards with resistive and random connections \cite{Nakata2013, Urade2015, Urade2016a, Takano2017}.
The checkerboard critical behavior is closely related to the checkerboard special property: the {\it local} electrical conductivity switching at the interconnection points
results in {\it global} Babinet inversion, which is the generalization of the interchange between the metallic and vacant areas.
We can interpret that local conductivity switching can induce a global {\it phase transition}, which exhibits
critical changes in the electromagnetic scattering property.
This transition is preferable to realizing not only the deep modulation
but also simultaneous design of dynamic metasurfaces,
because the original and switched states are related
through the structural symmetry of Babinet inversion.
This theory has been demonstrated to realize
the capacitive-inductive reconfigurable filter \cite{Urade2016}
and extended to anisotropic checkerboards to realize a reconfigurable polarizer \cite{Nakata2016}.
Even planar chirality switching has been theoretically proposed \cite{Urade2017}.

In this study, we propose a reconfigurable metasurface
based on the anisotropic checkerboard transition
to realize helicity switching, which requires
the deep modulation and simultaneous design of the switchable states.
As shown in Fig.~\ref{fig1}, this metasurface converts incident linear polarization
to right- or left-circular polarization, depending on the connection states.
The device functions as a dynamic quarter-wave plate,
the fast axis of which can be rotated by 90$^\circ$.
Compared to a previously studied dynamic polarizer
based on an anisotropic checkerboard \cite{Nakata2016},
of which only the amplitude manipulation function was investigated,
the dynamic quarter-wave plate requires the 
simultaneous control of both the amplitude and phase for each polarization.
We show that the anisotropic checkerboard
can also function as a dynamic quarter-wave plate because of a general
constraint between the transmission amplitude and phase,
and we demonstrate its functionality in the terahertz frequency range.
In this study, we assume a harmonic time dependence $\exp(\jj \omega t)$,
where we use $\jj$ for the imaginary unit.

\section{Principle}

\begin{figure}[!tbph]
\centering
\includegraphics[width=3.4in]{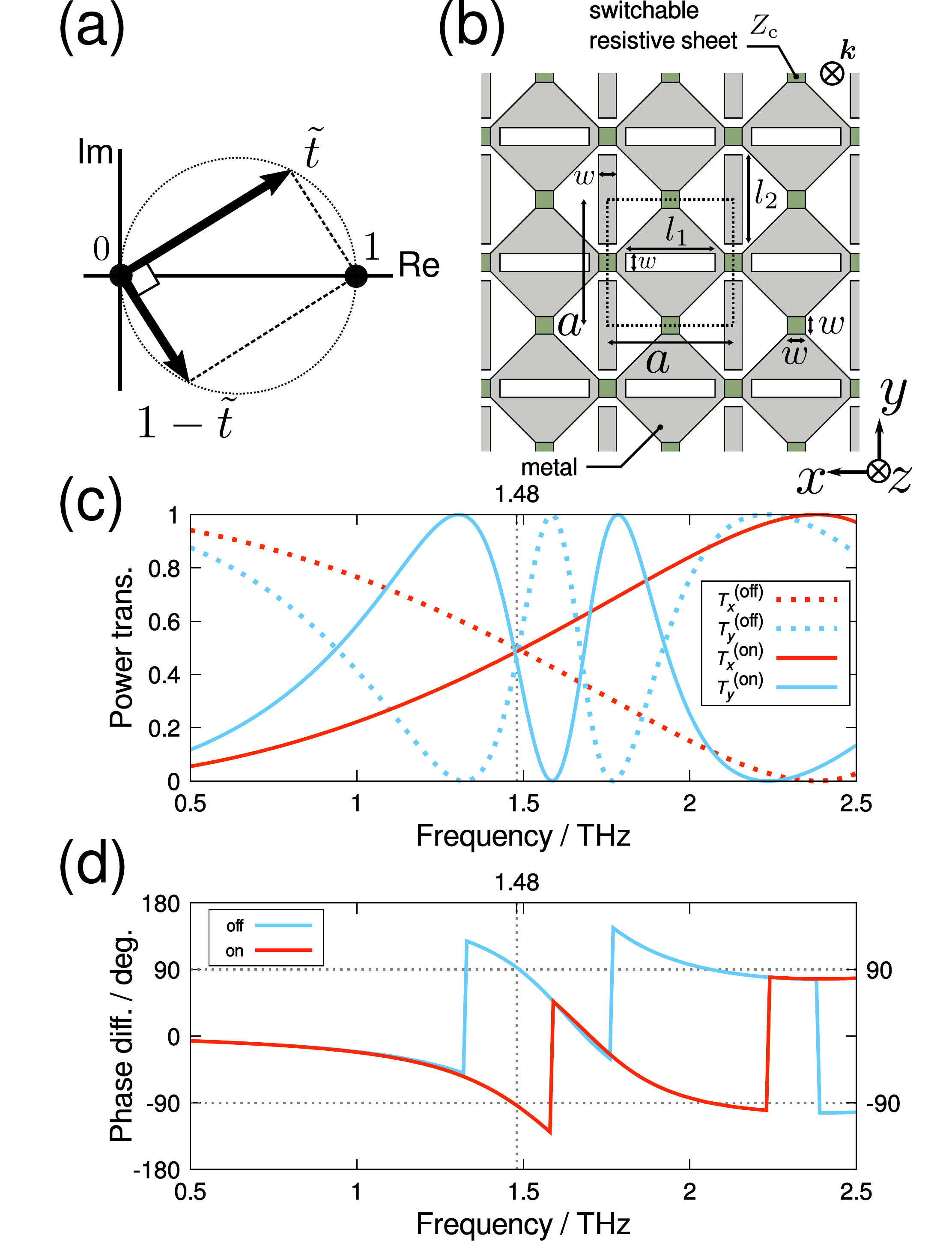}
 \caption{(a) Constraint on a complex transmission amplitude
for lossless electric metasurface without resistive elements, polarization conversion,
 or diffraction into higher-order modes. (b) Design of a dipole-nested checkerboard metasurface. 
Calculated (c) power transmission
($T_x =|\tilde{t}_x|^2$,  $T_y=|\tilde{t}_y|^2$)
 and
 (d) phase difference between transmitted $x$ and $y$-polarizations
[$\arg(\tilde{t}_y)-\arg(\tilde{t}_x)$]
 for the off- and on-state dipole-nested checkerboard made of perfect metal with
 $a=106\,\U{\upmu m}$, $w=15\,\U{\upmu m}$, $l=l_1=l_2=75\,\U{\upmu m}$
 at the normal incidence of plane waves.
 }
\label{fig2}
\end{figure}

First, we design an ideal dynamic quarter-wave plate
based on the anisotropic checkerboard
to demonstrate our strategy for realizing fast axis rotation.
Before going into detailed discussion,
we explain the general restriction imposed on the complex transmission amplitudes,
derived from energy conservation law.
Consider a thin electric metasurface located at $z=0$ in a vacuum.
An electromagnetic wave having a complex electric field 
$\tilde{\vct{E}}_0\exp(-\jj k_0 z)$ with $k_0>0$ enters the metasurface from $z<0$ to $z>0$.
The transmitted and reflected components propagating along the $z$-direction with the same polarization $\tilde{\vct{E}}_0$
are denoted by $\tilde{t}\tilde{\vct{E}}_0 \exp(-\jj k_0 z)$ and $\tilde{r}\tilde{\vct{E}}_0 \exp(\jj k_0 z)$, respectively.
Here, $\tilde{t}$ ($\tilde{r}$) is a complex transmission (reflection) coefficient.
If there is no energy loss caused by resistive elements, polarization conversion,
or diffraction into higher-order modes,
the following equation must be satisfied from
the electric field continuation $1+\tilde{r}= \tilde{t}$ at $z=0$ and
energy conservation law $|\tilde{t}|^2+|\tilde{r}|^2=1$:
\begin{equation}
 |\tilde{t}|^2+|1-\tilde{t}|^2=1.  \label{eq:1}
\end{equation}
This implies that $\tilde{t}$ must be on the circumference of the dashed circle
shown in Fig.~\ref{fig2}(a).
The quarter-wave plate function requires 
$\tilde{t}_y=\pm \jj \tilde{t}_x$,
where $\tilde{t}_i$ is $\tilde{t}$ for a normally incident $i$-polarized wave.
If we assume Eq.~(\ref{eq:1}) for each polarization, the
only possible choice for a quarter-wave plate is 
$\tilde{t}_x=(1+\jj)/2$, $\tilde{t}_y =(1-\jj)/2$,
or $\tilde{t}_x=(1-\jj)/2$, $\tilde{t}_y = (1+\jj)/2$ \cite{Baena2015, Baena2017}.
Under these conditions, $T_x = T_y =1/2$ is satisfied
for the power transmission $T_x = |\tilde{t}_x|^2$, $T_y = |\tilde{t}_y|^2$.

Now, we apply the checkerboard singularity
to quarter-wave plate switching.
Consider a dipole-nested checkerboard metasurface
(D-checkerboard)
at $z=0$ in a vacuum, as shown in Fig.~\ref{fig2}(b).
The thickness of the metal is safely assumed to be zero
if we consider the structure as much thinner than the typical length, such as $a$.
The metasurface is composed of perfectly metallic parts connected with
interconnection patches exhibiting variable sheet impedance $Z\sub{c}$.
Changing $Z\sub{c}$ from $\infty$ to $0$,
we can induce the checkerboard transition from
the disconnected structure (off state) to the connected structure (on state).
The metasurface is designed to be {\it dynamically Babinet-invertible} for $l=l_1=l_2$;
the on-state structure is obtained by $90^\circ$ rotation after Babinet inversion of the off-state structure.
This special symmetry automatically leads to the transmission
inversion for each polarization \cite{Nakata2016}:
\begin{equation}
\tilde{t}\sur{(off)}_i +  \tilde{t}\sur{(on)}_i =  1\hspace{2em} (i=x,y),  \label{eq:2}
\end{equation}
where $\tilde{t}\sur{(off)}_i$ and $\tilde{t}\sur{(on)}_i$ are $\tilde{t}_i$ for the
off and on state checkerboards, respectively.
If we assume Eq.~(\ref{eq:1}),
the transmission inversion
($\tilde{t}\sur{(off)}_i$ to $\tilde{t}\sur{(on)}_i = 1-\tilde{t}\sur{(off)}_i$)
is represented by $90^\circ$ rotation in the complex plane, as shown
in Fig.~\ref{fig2}(a).
If we have $\tilde{t}_x\sur{(off)}=(1+\jj)/2$ and $\tilde{t}\sur{(off)}_y =(1-\jj)/2$, then
$\tilde{t}_x\sur{(on)}=(1-\jj)/2$ and $\tilde{t}\sur{(on)}_y =(1+\jj)/2$
automatically follow, respectively.
Thus, we can realize dynamic rotation of
the fast axis of the quarter-wave plate by $90^\circ$.

To confirm the above theory,
we demonstrate an ideal reconfigurable quarter-wave
plate without a substrate, in a simulation.
To calculate the transmission spectra of the metasurface in a vacuum,
a conventional finite element solver (\textsc{COMSOL Multiphysics}) was used.
All the metal sections are assumed to be perfect electric conductors.
Plane waves with $x$- and $y$-polarizations normally enter into
a metasurface from $z<0$ to $z>0$.
Using mirror symmetry, the
simulation domain is set to a quarter
of the unit cell with appropriate
side boundary conditions depending on the incident polarization.
The top and bottom boundaries
are set as ports with only fundamental modes
because all calculations are performed
under the {\it diffraction frequency} $f\sub{d}=c_0/a=2.83\,\U{THz}$
($c_0$ is the speed of light in a vacuum),
above which waves can be diffracted into higher-order modes.
Fixing $a=106\,\U{\upmu m}$ and $w=15\,\U{\upmu m}$,
we search the condition to
realize $T_x =T_y=0.5$ by varying $l(=l_1 = l_2)$.
Performing parametric sweeps,
we found that $T_x=T_y = 0.5$ is realized at
$1.48\,\U{THz}$
for the off-state structure with $l=75\,\U{\upmu m}$.
Figure~\ref{fig2}(c) presents
the power transmission spectra for
both the off- and on-state structures with $l=75\,\U{\upmu m}$.
We can see that the four curves cross at $1.48\,\U{THz}$
with a power transmission of 0.5, as we predicted before.
Figure~\ref{fig2}(d) presents 
the phase difference $\arg(\tilde{t}_y) -\arg(\tilde{t}_x)$.
The phase difference is shifted from $+90^\circ$
to $-90^\circ$ at $1.48\,\U{THz}$ for the conductivity switching.
These characteristics are guaranteed by transmission inversion
because of the structural symmetry of the anisotropic checkerboard.
Thus, we could realize a subwavelength terahertz quarter-wave plate,
of which the fast axis can be dynamically rotated by 90$^\circ$.

\section{Experiment with optimized structure}
\begin{figure*}[!t]
 \centering
\includegraphics[width=6.4in]{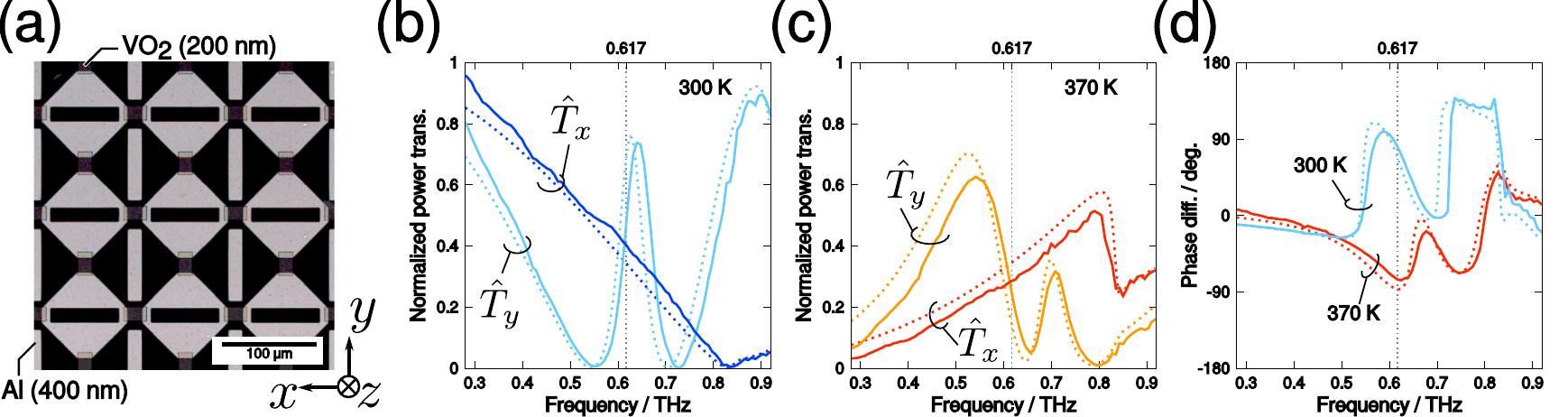}
 \caption{(a) Microphotograph of the fabricated sample with target lengths
 $a=106\,\U{\upmu m}$, $w=15\,\U{\upmu m}$, $l_1=72\,\U{\upmu m}$, and $l_2=76\,\U{\upmu m}$. Normalized power transmission spectra of the sample
 at (b) $300$ and (c) $370\,\U{K}$.
 (c) Phase difference between transmitted $x$ and $y$-polarizations for $300$ and $370\,\U{K}$. The simulated data (dotted) are also depicted with the experimental data (solid).
}
\label{fig3}
\end{figure*}

Although the above theoretical consideration works quite well,
a substrate is required for a practical device to hold the structures.
There is one way to use double-sided substrates sandwiching the structure
to assure Babinet inversion \cite{Urade2015}. However, we use another approach to slightly modify
the structure with a single-sided substrate for simple implementation.
In this case, Eq.~(\ref{eq:2}) barely holds,
but we can empirically realize quarter-wave plate switching
as follows.
In the experiments, structures are fabricated on a c-cut sapphire substrate.
A vanadium dioxide ($\mathrm{VO}_2$) film grown on the substrate is used as the variable resistance sheets.
It exhibits an insulator-to-metal transition of $T\sub{c}\approx
340\,\U{K}$, where the electrical conductivity
varies by several orders of magnitude through the transition.
To compensate for the substrate effect, we break the symmetry
condition $l_1=l_2$ and tune $l_1$ and $l_2$ individually in a simulation
to realize quarter-wave plate switching
(the other parameters are identical to the previous simulation).
The plane wave normally enters into the metasurface at $z=0$
from a vacuum ($z<0$) to the sapphire ($z\geq 0$)
with refractive index $n_x=n_y=n_ \perp\approx 3.1$ and
$n_z=n_\parallel\approx 3.4$ \cite{Grischkowsky1990}.
Perfectly matched layer-backed ports
are used to calculate spectra above the diffraction frequency
$f\sub{d} = c_0/(n_\parallel a)=0.83\,\U{THz}$.
Metallic parts are assumed to be aluminum with a
thickness of $400\,\U{nm}$ and conductivity $\sigma = 22\, \U{S/\upmu m}$ \cite{Laman2008}.
The variable resistive connection patch made of $\mathrm{VO}_2$
with a thickness of $200\,\U{nm}$ is represented
by the sheet impedance of $500\,\U{k\Omega}$ (off state)
and $10\,\U{\Omega}$ (on state), which are based on DC measurements.
The transition boundary condition with the background vacuum permittivity and permeability
is imposed onto both the Al and $\mathrm{VO}_2$ parts with
their conductivity and thickness.
We evaluate the normalized transmission coefficient of the metasurface as $\hat{t}_i=\tilde{t}_i/t_0$,
where $t_0=2/(1+n_\perp)$ is the Fresnel coefficient for transmission from a vacuum to the sapphire
\footnote{$\tilde{t}_i$ is also defined through the transmitted component $\tilde{t}_i \tilde{\vct{E}}_i \exp(-\jj n_\perp k_0 z)$ in $z\geq 0$
for the $i$-polarized incident wave $\tilde{\vct{E}}_i \exp(-\jj k_0 z)$ in $z<0$.}.
To achieve quarter-wave plate switching,
we tune $l_1$ and $l_2$ as much as possible
to realize $\hat{T}\sur{(off)}_x \approx \hat{T}\sur{(off)}_y \approx \hat{T}\sur{(on)}_x \approx \hat{T}\sur{(on)}_y$ at some frequency, where
$\hat{T}_i = |\hat{t}_i|^2$ and on/off represents the connection states.
This approximately leads to the switching function.
Through the optimization process from $l_1=l_2=75\,\U{\upmu m}$,
we reached $l_1=72\,\U{\upmu m}$ and $l_2=76\,\U{\upmu m}$.
The detailed simulation results are presented in Fig.~\ref{fig3}.

Using these target parameters, we fabricated a D-checkerboard on
a sapphire (0001) substrate (thickness: $1.0\,\U{mm}$).
A $\mathrm{VO}_2$ thickness of $200\,\U{nm}$ was sputtered
and patterned by photolithography using wet etching.
The D-checkerboard made of aluminum (thickness: $400\,\U{nm}$)
was formed by photolithography, electron beam
evaporation at room temperature, and the lift-off technique.
To ensure electrical connection, the $\mathrm{VO}_2$ squares
were partially overlapped by the Al structures.
Figure~\ref{fig3}(a) presents a top-view microphotograph
of the fabricated sample.
To evaluate the D-checkerboard, we performed polarization measurements using
conventional terahertz time-domain spectroscopy.
The sample holder was maintained at $300$ and $370\,\U{K}$
for the off and on states, respectively.
To extend the sample substrate to $3\,\U{mm}$
for delaying multiple reflection signals,
two pieces of $1\,\U{mm}$-thick c-cut sapphire substrates were placed under the sample.
As a reference sample,
three stacked pieces of the substrates were used.
Multiple reflection signals were truncated using
a rectangular time-domain window function.
Collimated terahertz waves normally enter into the metasurface.
Here, we define $\vct{e}_D=(\vct{e}_x+\vct{e}_y)/\sqrt{2}$ and
$\vct{e}_X=(-\vct{e}_x+\vct{e}_y)/\sqrt{2}$, where $\vct{e}_x$ and $\vct{e}_y$ are the unit vectors along the $x$ and $y$-axes, respectively.
Using wire grid polarizers, we measured $\hat{t}_{pq}$ $(p,q = D, X)$,
which is a normalized complex transmission amplitude from
a normally incident wave with $q$-polarization 
into a normally transmitted wave with $p$-polarization.
The normalization is performed through
$\tilde{E}_p(\omega)/\tilde{E}_q\sur{(ref)}(\omega)$, where $\tilde{E}_p(\omega)$ represents
the Fourier-transformed transmitted electric field with
polarization $p$ for the sample,
and $\tilde{E}_q\sur{(ref)}(\omega)$ is
that with polarization $q$ for the reference.
After the measurement, we convert $\hat{t}_{pq}$ $(p,q = D, X)$
into $\hat{t}_{ij}$ $(i,j =x,y)$ using matrix transformation.

Figures~\ref{fig3}(b) and (c) present the experimental (solid) and simulated (dotted) data of
normalized power transmission spectra
$\hat{T}_x=|\hat{t}_{x}|^2$ and $\hat{T}_y=|\hat{t}_{y}|^2$
for the off and on states.
The experimental data agree quite well with the simulated data.
The small discrepancy between the experimental and simulated data is
due to the inaccuracy of the fabrication.
At $0.617\,\U{THz}$, $\hat{T}_x\sur{(off)} \approx \hat{T}_y\sur{(off)}\approx \hat{T}_x\sur{(on)}\approx \hat{T}_y\sur{(on)}$ is satisfied.
In Fig.~\ref{fig3}(d), the phase difference
$\arg(\hat{t}_y) -\arg(\hat{t}_x)$ is depicted.
Phase difference switching from about $+90^\circ$ to $-90^\circ$ is observed
at $0.617\,\U{THz}$. This implies that we have realized 
fast axis rotation of the quarter-wave plate by $90^\circ$.
The maximum device thickness without the substrate is $600\,\U{nm}$,
which is deeply subwavelength at the working frequency.

\begin{figure}[!b]
\centering
\includegraphics[width=3in]{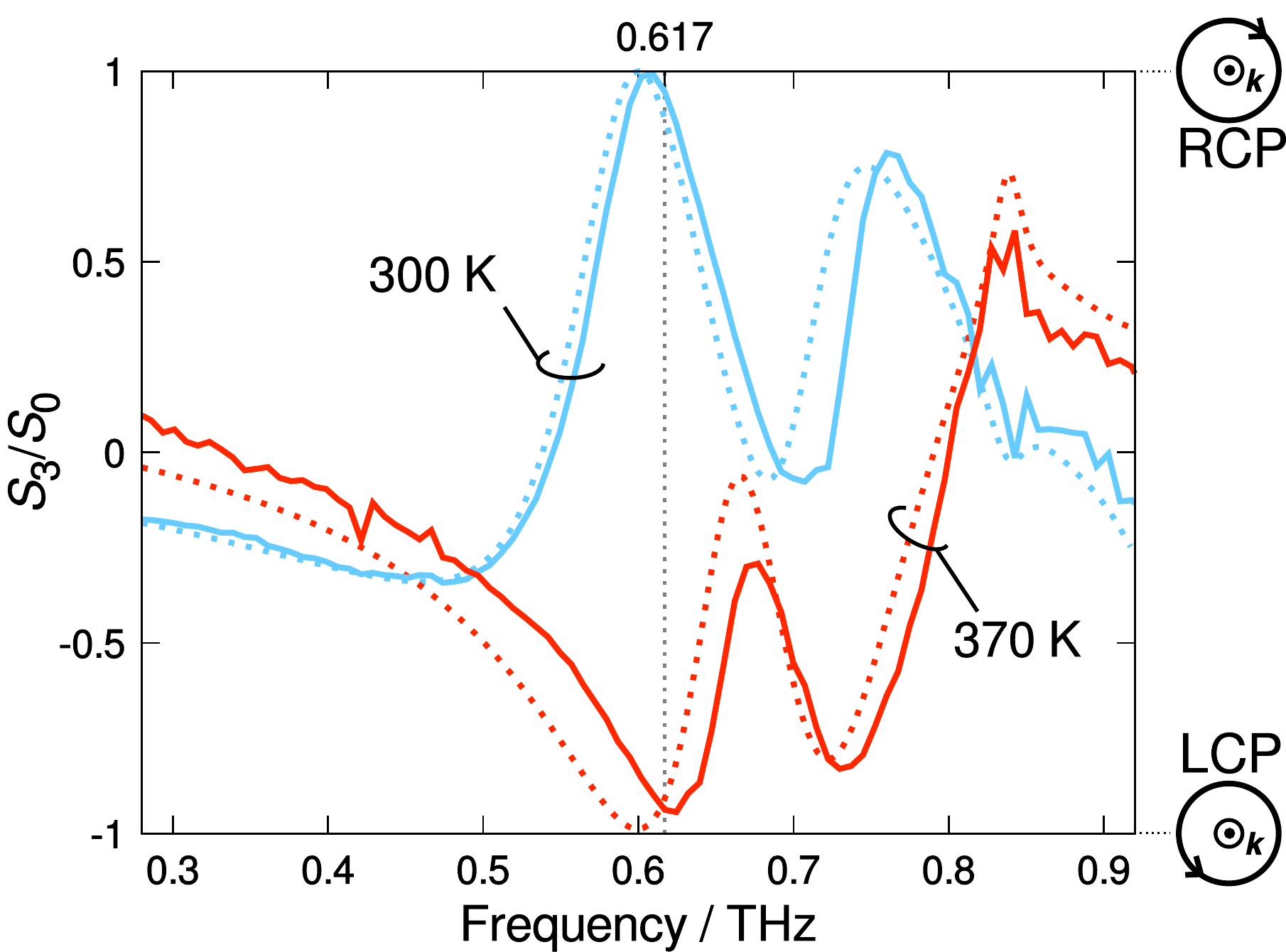}
  \caption{$S_3/S_0$ of the transmitted wave for
  a normally incident plane wave with $D$-polarization
  under $300$ and $370\,\U{K}$. The simulated data (dotted) are
  also depicted with the experimental data (solid).}
\label{fig4}
 \end{figure}

To apply this functionality to the helicity switching of a circular polarization,
the polarization of the transmitted wave was analyzed for
a normally incident wave with $D$-polarization.
Here, we evaluate $S_3/S_0$ with the Stokes parameters
$S_3$ and $S_0$ \cite{Saleh2007}, where $S_3/S_0=2 \Im (\hat{t}_{xD}^* \hat{t}_{yD})/(|\hat{t}_{xD}|^2+|\hat{t}_{yD}|^2)$ represents the $z$-coordinate
in the Poincar\'{e} sphere.
The right and left circularly polarized waves correspond to
$S_3/S_0=+1, -1$, respectively. 
Figure~\ref{fig4} presents the $S_3/S_0$ of the transmitted wave.
The helicity of the transmitted wave is dynamically switched at $0.617\,\U{THz}$
from $S_3/S_0 =0.95$ to $-0.94$ in the experiment.

\section{Conclusions}
We have proposed a terahertz metasurface functioning as a
reconfigurable quarter-wave plate, of which the fast axis can be dynamically rotated by $90^\circ$.
The device is based on the checkerboard critical transition, which
realizes the deep modulation and simultaneous design of
the off and on states of the device. 
Combining the constraint on complex transmission amplitudes
with transmission inversion,
we theoretically showed that $90^\circ$ fast axis rotation can be realized.
Performing simulations for the ideal structure without a substrate,
we confirmed that the theory works well.
To compensate the substrate effect for realizing practical experiments,
we adjusted the structural parameters
by slightly breaking the structural symmetry
on the interchange of the metallic and vacant area.
The optimized D-checkerboard was fabricated and evaluated
in the terahertz range. The experimental data agreed very well
with the simulated data, and the fast axis rotation of the quarter-wave plate was
realized at $0.617\,\U{THz}$.
By analyzing the polarization of the transmitted wave 
for a normally incident $D$-polarization wave,
we demonstrated the helicity switching function of the device.
This device is applicable to chirality-sensitive spectroscopy
for materials with circular dichroism and optical activity,
highly sensitive measurements, and data transmission
with circular polarization.
The demonstrated device was practically designed with
deeply subwavelength structures on thick substrates
to truncate multiple reflection signals.
For more efficient devices, we should further consider
the elimination of multiple reflections by using
a thinner substrate or anti-reflecting coating at the backside of the substrate.
The proposed design strategy to realize the reconfigurable quarter-wave plate
is not limited to the terahertz frequency range; it is applicable to other frequency
ranges, where variable resistance materials can be used.
By utilizing photo-carrier injection into semiconductors,
ultrafast polarization control can be achieved.
Broadband switching should also be studied in the future,
although subwavelength devices commonly require resonant behavior,
which leads to narrowband operation.

\section*{Acknowledgment}
The sample fabrication was performed with the help of the Kyoto University Nano Technology Hub, as part of the “Nanotechnology Platform Project,” sponsored by MEXT, Japan.
The present research is supported by a grant from the Murata Science Foundation and by JSPS KAKENHI Grant Nos. 17K17777 and 17K05075.

%\bibliographystyle{apsrev4-1_custom}
%\bibliography{main}


\end{document}